\documentclass[aps,prd,amsmath,amssymb,preprintnumbers,nofootinbib,a4paper,11pt]{revtex4}
\pdfoutput=1
\usepackage{amsthm}
\usepackage{graphicx}
	\graphicspath{{./NewFig/}}
\usepackage{url}
\usepackage[bookmarks, pagebackref=false]{hyperref}
\usepackage{color}
	\definecolor{rossoCP3}{cmyk}{0,.88,.77,.40}
		\definecolor{graa}{rgb}{0.8,0.8,0.8}
		\definecolor{blaa}{rgb}{0.2,0.2,0.6}
		\hypersetup{
			colorlinks, 
			bookmarksopen, 
			bookmarksnumbered,
			citecolor=blaa, 		
			linkcolor=rossoCP3,	
			urlcolor=rossoCP3,			
			}
\usepackage{dcolumn}
\usepackage{bm}
\usepackage{bbm}
\usepackage{pxfonts}

\usepackage{epsfig}
\usepackage{placeins}

\usepackage{bbold}

\usepackage[ margin=5pt, font=small,labelfont=bf, justification=raggedright]{caption}
\usepackage{youngtab}
\usepackage{slashed}
\usepackage{subfig}

\newcommand{\beq}{\begin{eqnarray}}
\newcommand{\eeq}{\end{eqnarray}}

\newcommand{\bmp}{\noindent\begin{minipage}{16cm}}
\newcommand{\emp}{\end{minipage}\vskip 7mm} 

\def\lsim{\mathrel{\rlap{\lower4pt\hbox{\hskip1pt$\sim$}}
    \raise1pt\hbox{$<$}}}                
\def\gsim{\mathrel{\rlap{\lower4pt\hbox{\hskip1pt$\sim$}}
    \raise1pt\hbox{$>$}}}                

\begin{document}

\title{\LARGE \color{rossoCP3} Vacuum Alignment with more Flavors}
 \author{Thomas A. Ryttov}\email{ryttov@cp3.dias.sdu.dk} 
  \affiliation{
{ \color{rossoCP3}  \rm CP}$^{\color{rossoCP3} \bf 3}${\color{rossoCP3}\rm-Origins} \& the Danish Institute for Advanced Study {\color{rossoCP3} \rm DIAS},\\ 
University of Southern Denmark, Campusvej 55, DK-5230 Odense M, Denmark.
}

\begin{abstract}
We study the alignment of the vacuum in gauge theories with $N_f$ Dirac fermions transforming according to a complex representation of the gauge group. The alignment of the vacuum is produced by adding a small mass perturbation to the theory. We study in detail the $N_f=2,3$ and $4$ case. For $N_f=2$ and $N_f=3$ we reproduce earlier known results including the Dashen phase with spontaneous violation of the combined charge conjugation and parity symmetry, CP. For $N_f=4$ we find regions with and without spontaneous CP violation.

We then generalize to an arbitrary number of flavors. Here it is shown that at the point where $N_f-1$ flavors are degenerate with positive mass $m>0$ and the mass of the $N_f$'th flavor becomes negative and equal to $-m$ CP breaks spontaneously. 
 \vskip .1cm
{\footnotesize  \it Preprint:  CP$^3$-Origins-2014-017 DNRF 90\ \& DIAS-2014-17}
 \end{abstract}

\maketitle

\newpage
     
\section{Introduction}

Despite the fact that QCD has been with us since the discovery of asymptotic freedom \cite{Gross:1973id,Gross:1973ju,Gross:1974cs} around 1973 there are still subtle issues that hunger for an explanation. This includes the mechanism of color confinement and chiral symmetry breaking. Color confinement causes the quarks to be bound inside hadrons while chiral symmetry breaking is the source of the lightness of the pions being the associated (pseudo) Nambu-Goldstone bosons \cite{Nambu:1961tp,Nambu:1961fr} of the spontaneous symmetry breaking.  

A few nonperturbative and exact results are known about chiral symmetry breaking. For instance Coleman and Witten have shown that under reasonable assumptions the chiral symmetry group must break down to its diagonal subgroup in the limit of a large number of colors \cite{Coleman:1980mx}. Another important result has been derived by Vafa and Witten who showed that in parity-conserving theories (such as QCD) parity cannot be spontaneously broken \cite{Vafa:1984xg}. 

Understanding the dynamics of massless QCD is of course of vital importance but the massiveness of the pions forces us to study the system including small perturbations. The fact that the pions are light (but nonzero) is due to the quarks requiring an explicit mass thereby explicitly breaking the chiral symmetry. The explicit symmetry breaking causes the vacuum degeneracy to be lifted and the associated Nambu-Goldstone bosons become massive.

One can imagine that this small mass perturbation drives the vacuum of the system in a new direction, i.e. the vacuum of the theory will align with the perturbation. In which direction the vacuum will align will depend on the details and nature of the perturbation. The study of this vacuum alignment phenomena in QCD was first done by Dashen \cite{Dashen} who found a very interesting result. For three quarks (up, down and strange) where all three quark masses are negative, the up and down quark masses are degenerate and the numerical value of the up and down quark masses are not too big compared to the numerical value of the strange quark mass the vacuum aligns in a direction in which the combined symmetry of charge conjugation and parity, CP, is spontaneously broken  \cite{Dashen}. 

Since the scale of this CP violation would be the scale of chiral symmetry breaking it is clear that this cannot be the source of the observed CP violation in Nature. Some time after the work of Dashen it was suggested by Eichten, Lane and Preskill that within the technicolor framework of physics beyond the Standard Model a similar mechanism could account for the observed CP violation \cite{Eichten:1980du}. Additional work can be found in \cite{Martin:2004ec,Lane:2005we,Lane:2005vp,Lane:2000es,Rador:2009sy,Lane:2002wv}.  In this scenario the procedure for generating the observed CP violation in the Standard Model is very similar to the one originally performed by Dashen \cite{Dashen}. Hence besides being a fascinating phenomena within QCD vacuum alignment has also found its relevance in other branches of high energy physics. 

In QCD for two and three flavors it has been studied in detail by M. Creutz and collaborators in a series of papers \cite{Creutz:1995wf,Creutz:2003xu,Creutz:2003xc,Creutz:2005gb,Creutz:2013xfa,Aoki:2014moa,Creutz:2000bs}.  
For a larger number of flavors there has been little work done.  It is the purpose of this work to study the alignment of the vacuum when the matter sector consists of four flavors or higher. As already mentioned one of the most intriguing observations for three flavors is the fact that there is a region in parameter space where CP is spontaneously broken. This is distinct from the two flavor case where only a single point in the parameter space (where the up quark mass is equal to minus the down quark mass) can yield spontaneous CP violation. Whether this phenomenon continues to occur for a larger number of flavors is still an open question. As will be clear in the following when minimizing the energy of the system the existence of a nontrivial solution (where CP is broken) depends highly on the number of flavors. 

We will see at the point where $N_f-1$ flavors are degenerate with positive masses $m>0$ and the mass of the $N_f$'th flavor becomes negative and equal to $-m$ CP breaks spontaneously for both three and four flavors. We are then able to show that this happens for any number of flavors. In other words it is sufficient for CP to be violated that only the mass of a single flavor becomes negative and equal to $-m$ for any number of flavors. As one varies the number of flavors the phases with spontaneous CP violation are smoothly connected.

The paper is organized as follows: In Section \ref{sec:vacalignment} we discuss vacuum alignment in generality and in Section \ref{sec:symmetries} we discuss the global and discrete symmetries that may or may not be broken. In Section \ref{sec:results} we review the case of two and three flavors and present new results for four flavors and an arbitrary number of flavors. We finally conclude in Section \ref{sec:conclusion}.

\section{Vacuum Alignment}\label{sec:vacalignment}

We begin our discussion of vacuum alignment following and generalizing the original work of Dashen \cite{Dashen}. For now we shall not commit ourselves to a specific theory but consider generic asymptotically free nonsupersymmetric gauge theories with a set of massless fermions belonging to some representation of the gauge group. The Hamiltonian of the theory is denoted as $\mathcal{H}_0$ and we shall assume that it possesses a global symmetry $G$. In general the global symmetry depends on the representation to which the fermions belong \cite{Peskin:1980gc}. We also imagine picking the number of flavors to be less than the critical value needed to reach an infrared fixed point \cite{Caswell:1974gg,Banks:1981nn,Appelquist:1988yc,Appelquist:1997dc,Sannino:2004qp,Dietrich:2006cm,Ryttov:2009yw,Ryttov:2010hs,Mojaza:2012zd,Antipin:2013qya,Ryttov:2007cx,Pica:2010mt,Ryttov:2013ura,Ryttov:2013hka,Ryttov:2012qu,Pica:2010xq,Ryttov:2007sr,Shrock:2013cca,Molgaard:2014mqa,Ryttov:2010iz}. 

Then at a certain energy scale the gauge interactions become strong and spontaneously break the global symmetry $G$ to a subgroup $G'$. Even though the Hamiltonian $\mathcal{H}_0$ is invariant under $G$ the ground state of the theory is invariant only under $G'$
\begin{eqnarray}
U(g') | \Omega \rangle &=&  | \Omega \rangle \ , \qquad \text{if $g' \in G'$} \ , \\
U(w) | \Omega \rangle & \neq & | \Omega \rangle \ , \qquad \text{if $w \in G/G'$}
\end{eqnarray}
where $U(g')$ and $U(w)$ are unitary representations of the group elements $g' \in G'$ and $w \in G/G'$ respectively. Similar to the fact that $|\Omega \rangle$ is a ground state of the theory so is $U(w) | \Omega \rangle$ since they both have the same energy
\begin{eqnarray}\label{eq:equalenergy}
\langle \Omega | U^{\dagger}(w) \mathcal{H}_0 U(w) | \Omega \rangle
= \langle \Omega | \mathcal{H}_0 | \Omega \rangle \ .
\end{eqnarray}
Here we have used the fact that the Hamiltonian is invariant under all group transformations and therefore specifically under transformations $w \in G/G'$. The ground state is therefore not unique. Instead the theory contains an infinite number of vacua $U(w)| \Omega \rangle \equiv | \Omega_w \rangle$ parameterized by all the transformations $w \in G/G'$. One should note that there is nothing special about the ground state $| \Omega \rangle = | \Omega_{w_0} \rangle$ compared to the others. It is just to be understood as some standard reference vacuum corresponding to a particular transformation $w_0$.

It is a well known fact that the above vacuum degeneracy is manifested in the presence of a set of massless Nambu-Goldstone bosons corresponding to the number of broken generators of the global symmetry. However out of the infinite number of vacua is there any way that we can determine which one to pick and from which we should build our particle states, etc.?

Let us now imagine that to the Hamiltonian $\mathcal{H}_0$ we add a small perturbation $\mathcal{H}'$ which explicitly breaks the global symmetry $G$. The explicit symmetry breaking will cause the vacuum degeneracy to be either partially or completely lifted. The vacuum (vacua) that remains in the theory is (are) said to be aligned with the explicit symmetry breaking term. 

To find the correct (aligned) vacuum we need to minimize the energy
\begin{eqnarray}
E &=& \langle \Omega_w | \left(  \mathcal{H}_0 + \mathcal{H}'   \right)  | \Omega_w \rangle \ , \qquad
\end{eqnarray}
over all vacua of the unperturbed theory. As discussed above the first term is just a constant and independent of the specific vacua we consider. Therefore when finding the correct (aligned) vacuum we only need to consider the minimization of the second term (which we shall also denote by $E$. This should not cause any confusion.)
\begin{eqnarray}
 E &=&  \langle \Omega_w | \mathcal{H}' | \Omega_w \rangle
\end{eqnarray}
 
So far we have discussed vacuum alignment in the passive picture in which we keep the Hamiltonian fixed and act on the set of degenerate ground states by unitary transformations. In this way one finds the aligned vacuum state compatible with the Hamiltonian. On the other hand one could equally well fix the ground state and then act on the Hamiltonian by unitary transformations. This is the active picture. Switching from the passive to the active picture is straightforward. First let us write the part of the total energy we want to minimize as
\begin{eqnarray}
 E &=& \langle \Omega | U^{\dagger}(w) \mathcal{H}' U(w) | \Omega \rangle = \langle \Omega |  \mathcal{H}'_w  | \Omega \rangle
\end{eqnarray}
The above procedure of finding the vacuum $| \Omega_w \rangle$ that minimizes the energy with respect to some fixed Hamiltonian $\mathcal{H}'$ now translates into finding the rotated Hamiltonian $\mathcal{H}'_w$ that minimizes the energy with respect to some fixed ground state $| \Omega \rangle$ where
\begin{eqnarray}
\mathcal{H}'_w &=& U^{\dagger}(w) \mathcal{H}' U(w)
\end{eqnarray}
For the remaining part of this work we shall use both the passive and active picture and choose the one which is most convenient for us in order to illustrate various points and issues.

\section{Symmetries}\label{sec:symmetries}

Fermionic gauge theories enjoy global symmetries $G$ depending on the representation to which the fermions belong \cite{Peskin:1980gc}. If the representation is complex the theory possesses a continuous global symmetry $G=SU(N_f)_L \times SU(N_f)_R \times U(1)_B$ where $N_f$ counts the number of Dirac fermions. Also $U(1)_B$ is anomaly free and is the standard baryon number. 

In addition to the above continuous global symmetries a theory might also possess a set of discrete symmetries. Of particular interest is parity $P$ which exchanges left-handedness and right-handedness, charge conjugation $C$ which transforms a particle into its antiparticle and time reversal $T$. 

In order to discuss their implementation on various fields let us write a four-component Dirac fermion as
\begin{eqnarray}
\Psi_D &=& \left( \begin{array}{c}
\psi_L \\
\psi_R
\end{array} \right) \ ,
\end{eqnarray}
where $\psi_L$ and $\psi_R$ are the two left and right handed Weyl components. Under parity, charge conjugation and time reversal they transform as
\begin{eqnarray}
 \left( \begin{array}{c}
\psi_L \\
\psi_R
\end{array} \right)
& \stackrel{P}{\rightarrow} &
 \eta \left( \begin{array}{c}
\psi_R \\
\psi_L
\end{array} \right) \ , \qquad \eta\eta^*=1 \\
 \left( \begin{array}{c}
\psi_L \\
\psi_R
\end{array} \right)
& \stackrel{C}{\rightarrow} &
 \left( \begin{array}{c}
-i \sigma^2 \psi_R^* \\
i \sigma^2 \psi_L^*
\end{array} \right) \ , \\
 \left( \begin{array}{c}
\psi_L \\
\psi_R
\end{array} \right)
& \stackrel{T}{\rightarrow} &
 \left( \begin{array}{c}
- \sigma^1 \sigma^3 \psi_L \\
- \sigma^1 \sigma^3 \psi_R
\end{array} \right) \ ,
\end{eqnarray}
where $\eta$ is an arbitrary phase.\footnote{Time reversal $T$ is an antiunitary operator} An important property of Lorentz invariant local quantum field theories with a Hermitian Hamiltonian is that they preserve the combined $CPT$ symmetry. This is the celebrated $CPT$-theorem \cite{Luders,LudersZumino,Pauli}. For instance it implies that if $CP$ is preserved (violated) then $T$ is also preserved (violated). Invoking the $CPT$-theorem it therefore suffices to study only the possible realizations of parity $P$, charge conjugation $C$ and the combined $CP$ symmetry. This exhausts all the possible combinations.

At last we note that the gauge fields transform as
\begin{eqnarray}
A_{\mu}^a T^a & \stackrel{P}{\rightarrow} &  (-1)^{\mu} A_{\mu}^a T^a \ , \\
A_{\mu}^a T^a & \stackrel{C}{\rightarrow} & - A_{\mu}^a T^{a*} \ , \\
A_{\mu}^a T^a & \stackrel{T}{\rightarrow} & (-1)^{\mu} A_{\mu}^a T^a
\end{eqnarray}
under $P$ and $C$. Here $(-1)^{\mu}=1$ for $\mu=0$, $(-1)^{\mu}=-1$ for $\mu=1,2,3$ and there is no summation over $\mu$. Also the $T^a$ are the generators of the gauge group while $-T^{a*}$ generate the conjugate representation. For the readers convenience we give in Appendix \ref{App:Discrete} a list of how different bilinear operators transform under $C$, $P$ and $CP$.

\section{Results}\label{sec:results}

The theories we consider have a gauge symmetry with a set of $N_f$ Dirac fermions belonging to an arbitrary complex representation of the gauge group. We denote the Dirac fermions as
\begin{eqnarray}
\psi_L^f  \ , \ \psi_R^{f'} \ , \qquad f,f' = 1,\ldots,N_f \ .
\end{eqnarray}
The fermions also carry a gauge index which we have suppressed. The Lagrangian of the theory is
\begin{eqnarray}
\mathcal{L}_0 &=& -\frac{1}{4}  F_{\mu\nu}^a F^{a,\mu\nu}  + i \bar{\psi}_{L} \bar{\sigma}^{\mu} D_{\mu} \psi_L + i \bar{\psi}_{R} \sigma^{\mu} D_{\mu} \psi_R \ , \\
F_{\mu\nu}^a &=& \partial_{\mu} A_{\nu}^a - \partial_{\nu} A_{\mu}^a + g f^{abc} A_{\mu}^b A_{\nu}^c \\
D_{\mu} \psi_{L,R} &=& \partial_{\mu} \psi_{L,R} - i g A_{\mu}^a T^a_{r} \psi_{L,R}  \ , \\
a,b,c &=& 1,\dots,d_A
\end{eqnarray}
where $g$ denotes the gauge coupling and $d_A$ denotes the dimension of the adjoint representation of the gauge group. From the Lagrangian $\mathcal{L}_0$ we can easily construct the Hamiltonian $\mathcal{H}_0$. The theory enjoys a continuous global symmetry $G = SU(N_f)_L \times SU(N_f)_R \times U(1)_B $ where the abelian $U(1)_B$ is anomaly free and is identified with the standard baryon number. In addition to the continuous symmetry the theory is also separately invariant under parity $P$, charge conjugation $C$ and time reversal $T$. At last we note that if one picks the gauge group to be $SU(3)$ and the representation $r$ to be the fundamental representation the theory is simply ordinary massless QCD with $N_f$ flavors.

Next we imagine that the gauge interactions become strong at a certain energy scale triggering the formation of the condensate
\begin{eqnarray}\label{ComplexRep-Condensate}
\langle \Omega  | \bar{\psi}_{R,f} \psi_L^{f'} | \Omega \rangle &=&  - \frac{1}{2} \Delta \delta^{f'}_{\phantom{f}f} \ .
\end{eqnarray}
where we have absorbed a possible complex phase into the fields such that $\Delta >0$. The condensate breaks the continuous global symmetry $G$ to its vectorial subgroup $G'=SU(N_f)_V \times U(1)_B$ and is the source for the appearance of $N_f^2-1$ Nambu-Goldstone bosons. It should be emphasized that we have taken our reference vacuum to be the identity in which the condensate is particularly simple and CP is preserved.

Adding a perturbation that explicitly breaks the global symmetry $G$ the ground state degeneracy is lifted. As our example we will consider perturbing the theory by a small mass term
\begin{eqnarray}
\mathcal{H}' & = & \bar{\psi}_{R} M \psi_L + \bar{\psi}_L M^{\dagger} \psi_R
\end{eqnarray}
Here the fermion mass matrix $M$ is assumed to be real and diagonal. Depending on its eigenvalues it breaks the continuous global symmetry $G$ explicitly. However since the mass matrix is assumed to be real the perturbation is invariant under the discrete CP symmetry.

Having added a small perturbation are we sure that the identity is still the correct vacuum? As outlined above we want to minimize the energy
\begin{eqnarray}\label{complexenergy}
E(V) &=& \langle \Omega  | \mathcal{H}' _w| \Omega \rangle = - \frac{1}{2} \Delta \text{Tr}\left[ M V + M V^{\dagger} \right]
\end{eqnarray}
over all rotated vacua $V=g_Lg_R^{\dagger}$ where $g_{L,R} \in SU(N_f)_{L,R}$. Above we have used the fact that the mass matrix is real and diagonal. Also the rotated Hamiltonian is 
\begin{eqnarray}
 \mathcal{H}'_w &=& \bar{\psi}_R g_R^{\dagger} M g_L \psi_{L} + \bar{\psi}_L g_L^{\dagger} M^{\dagger} g_R \psi_{R}
\end{eqnarray}
It should be clear that if the (correct) aligned vacuum $V_{min}$ that minimizes the energy is complex then CP is spontaneously broken in this vacuum. Below we present the results.

\subsection{Two Flavors. $N_f=2$.}

We take the mass perturbation to be
\begin{eqnarray}
M &=& 
 \left(
\begin{array}{cc}
m_1 & 0 \\
0  & m_2
\end{array}
\right)
\end{eqnarray}
where both eigenvalues $m_1$ and $m_2$  are real such that CP is preserved. The perturbation breaks the global symmetry $G=SU(2)_L \times SU(2)_R \times U(1)_B$ explicitly.

The next step is then to find the aligned vacuum by minimizing Eq. \ref{complexenergy}. First, since $M$ is diagonal we can take $V=\text{diag}(v_1,v_2)$ to be diagonal as well. Second, since $V$ must be unitary both $v_1=e^{i\theta_1}$ and $v_2=e^{i\theta_2}$ must be pure phases. In addition since the determinant of $V$ must be equal to one this demands that $\theta_1 + \theta_2 = 2\pi n$ where $n$ is some integer. This last condition we shall enforce by a Lagrange multiplier. The aligned vacuum must therefore be of the form
\begin{eqnarray}
V &=& \left(
\begin{array}{cc}
e^{i\theta_1} & 0 \\
0 & e^{i\theta_2}
\end{array}
\right) \ , \qquad \qquad \theta_1 + \theta_2 = 2\pi n 
\end{eqnarray}
with
\begin{eqnarray}\label{eq:complex2}
 E &=& -\Delta \left( m_1 \cos \theta_1 + m_2 \cos \theta_2 \right)
\end{eqnarray}
We have to minimize the energy over the two angles $\theta_1$ and $\theta_2$ under the constraint that they must add to $2\pi n$. The extrema conditions follow quite simply and reads
\begin{eqnarray}
m_1 \sin \theta_1 &=& m_2 \sin \theta_2 \\
\theta_1 + \theta_2 &=& 2\pi n
\end{eqnarray}
Hence we are looking for solutions to $\left( m_1 + m_2\right) \sin \theta_1 = 0$. If $m_1 +m_2 \neq 0 $ they are located at $\theta_1 = \pi k$. Then if $m_1 + m_2 >0 $ they are global minima of the energy for even $k$ and global maxima of the energy for odd $k$. On the other hand if $m_1+m_2 <0$ they are global minima of the energy for odd $k$ and global maxima of the energy for even $k$. Lastly if $m_1 +m_2 = 0$ there exists a continuum of vacua all parameterized by an angle $\theta = \theta_1 = 2\pi n -\theta_2$ and the energy vanishes identically. Hence the correct aligned vacuum is
\begin{eqnarray}
V_{min} &=&\text{sgn}\left( m_1 + m_2 \right) \left(
\begin{array}{cc}
1 & 0 \\
0 & 1
\end{array}
\right) \ , \qquad \qquad m_1 +m_2  \neq 0 \\
V_{min} &=&  \left(
\begin{array}{cc}
e^{i\theta} & 0 \\
0 & e^{-i\theta}
\end{array}
\right) \ , \qquad \qquad \qquad \qquad \ \ m_1 + m_2 = 0 \label{eq:twoflavorsdegeneracy}
\end{eqnarray}
where $\text{sgn}(x)$ is the sign function. It equals $+1$ if $x$ is positive and $-1$ if $x$ is negative. Hence for $m_1 + m_2 \neq 0$ the vacuum degeneracy is lifted and only a single vacuum proportional to the identity remains. In other words all three Nambu-Goldstone bosons have become massive with masses proportional to $m_1 + m_2$. Now as we vary the quark masses and reach the point where $m_1+m_2 =0 $ the degeneracy appears again being parameterized by an angle $\theta$. This should not come as a big surprise since at this point the Nambu-Goldstone bosons will again appear massless.

We also note one interesting possibility discussed in a series of papers \cite{Creutz:1995wf,Creutz:2003xu,Creutz:2003xc,Creutz:2005gb,Creutz:2013xfa,Aoki:2014moa}. It is clear that at the point $m_1+m_2=0$ with $\theta \neq 0$ and $\theta \neq \pi$ the discrete CP symmetry is broken. Will the system somehow choose one of these vacua? In \cite{Creutz:1995wf,Creutz:2003xu,Creutz:2003xc,Creutz:2005gb,Creutz:2013xfa,Aoki:2014moa} it has been argued that at this point the neutral pion field becomes massless and condenses forming a spontaneous CP violating phase. We shall not speculate more on this possibility.

\subsection{Three Flavors. $N_f=3$.}

Let us now review the case of $N_f=3$ Dirac flavors with a mass perturbation
\begin{eqnarray}
M &=& 
\left( 
\begin{array}{ccc}
m_1 & 0 & 0 \\
0 & m_2 & 0 \\
0 & 0 & m_3 
\end{array}
\right)
\end{eqnarray}
where all eigenvalues are taken to be real such it preserves CP. The perturbation breaks the global symmetry explicitly where the pattern of symmetry breaking depends on the nature of the eigenvalues. Having added this perturbation we must again minimize Eq. \ref{complexenergy} in order to find the aligned vacuum. The argument is similar to the two flavor case. 

Since $M$ is diagonal we can take $V=\text{diag}(v_1,v_2,v_3)$ to be diagonal as well. Also $v_1=e^{i\theta_1}$, $v_2=e^{i\theta_2}$ and $v_3=e^{i\theta_3}$ must all be pure phases since $V$ is unitary. Lastly since the determinant of $V$ must be equal to one we have $\theta_1 + \theta_2+\theta_3 = 2\pi n$ where $n$ is some integer. This last condition should be enforced by a Lagrange multiplier. The aligned vacuum must therefore be of the form
\begin{eqnarray}
V &=& 
\left(
\begin{array}{ccc}
e^{i\theta_1} & 0 & 0  \\
0 & e^{i\theta_2} & 0 \\
0& 0 & e^{i\theta_3 }
\end{array}
\right) \ , \qquad \qquad \theta_1 + \theta_2 + \theta_3 = 2\pi n
\end{eqnarray}
with
\begin{eqnarray}\label{eq:complex3}
 E  &=& - \Delta \left( m_1 \cos \theta_1 + m_2 \cos \theta_2 + m_3 \cos \theta_3 \right)
\end{eqnarray}
Compared to two flavors the three flavor case is just a bit more involved. We need to minimize the energy over the three angles $\theta_1$, $\theta_2$ and $\theta_3$ subject to the constraint that they must add to $2\pi n$. It follows that the vacuum (extremum) conditions are
\begin{eqnarray}
m_1 \sin \theta_1 &=& m_2 \sin \theta_2 = m_3 \sin \theta_3 \ , \qquad \theta_1 + \theta_2 + \theta_3 = 2\pi n
\end{eqnarray}
It should be noted that they are identical to the ones obtained in \cite{Creutz:2003xu}. Let us solve this coupled set of equations in a number of different cases.

\begin{itemize}
\item $\text{sgn} (\det M)=1$: If all masses are positive then the energy is minimized when each cosine factor equals $+1$. Then each angle must be an even multiple of $\pi$ making the sum of the angles also an even multiple of $\pi$. The aligned vacuum therefore is
\begin{eqnarray}
V_{min} &=& \left( \begin{array}{ccc}
1 & 0 & 0 \\
0 & 1 & 0 \\
0 & 0 & 1
\end{array}
\right) \ , \qquad m_1,\ m_2,\ m_3>0
\end{eqnarray} 
On the other hand if only a single mass $m_i>0$ is positive while the remaining two $m_j,\ m_k <0$ are negative the energy is minimized when $\cos \theta_i = +1$ and $\cos \theta_j = \cos \theta_k = -1$. This is possible if $\theta_i$ is an even multiple of $\pi$ and the remaining two angles $\theta_j$ and $\theta_k$ are  odd multiples of $\pi$. It then follows that the sum of the angles add to an even number of $\pi$ as required. The aligned vacuum therefore is
\begin{eqnarray}
V_{min} &=& 
\left( 
\begin{array}{ccc}
1 & 0 & 0 \\
0 & -1 & 0 \\
0 & 0 & -1
\end{array}
\right) \ , \qquad m_1 >0 \ , \qquad m_2,\ m_3 <0 \\
V_{min} &=& 
\left( 
\begin{array}{ccc}
-1 & 0 & 0 \\
0 & 1 & 0 \\
0 & 0 & -1
\end{array}
\right) \ , \qquad m_2 >0 \ , \qquad m_1,\ m_3 <0 \\
V_{min} &=& 
\left( 
\begin{array}{ccc}
-1 & 0 & 0 \\
0 & -1 & 0 \\
0 & 0 & 1
\end{array}
\right) \ , \qquad m_3 >0 \ , \qquad m_1,\ m_2 <0
\end{eqnarray}
\end{itemize}
It is clear that the same simple line of arguments does not hold if all the masses are negative. Assuming that the energy is minimized when all the cosine factors are equal to $-1$ implies that the each angle must be equal to an odd multiple of $\pi$. However then they do not add up to an even multiple of $\pi$ as they should. A similar argument can be given if only a single mass is negative while the remaining two are positive.

In these situations we should therefore perform a more complete analysis of the vacuum conditions. Eliminating $\theta_3$ the vacuum conditions can be written as
\begin{eqnarray}
m_1 \sin \theta_1 = m_2 \sin \theta_2 \ , \qquad \sin \theta_2 \left( m_2 + m_3 \cos \theta_1 +  \frac{m_2m_3}{m_1} \cos \theta_2 \right) =0 
\end{eqnarray}
which can be satisfied in two different ways. Either
\begin{eqnarray}
 \sin \theta_1 &=& \sin \theta_2  = 0
\end{eqnarray}
or 
\begin{eqnarray}
m_1 \sin\theta_1 &=& m_2 \sin \theta_2 \\
0 &=& m_2 + m_3 \cos \theta_1 + \frac{m_2m_3}{m_1} \cos \theta_2
\end{eqnarray}
Let us consider the first scenario. Here all three angles must be equal to a multiple of $\pi$. This leaves us with four possibilities: Either all three angles are an even multiple of $\pi$, the first angle is an even multiple of $\pi$ with the remaining two being odd multiples of $\pi$, the second angle is an even multiple of $\pi$ with the remaining two being odd multiples of $\pi$ or the third angle is an even multiple of $\pi$ with the remaining two being odd multiples of $\pi$. At these four locations the value of the energy is 
\begin{eqnarray}
\label{eq:vac0}
 E_1 &=& -\Delta \left( m_1 + m_2 + m_3 \right)  \\
 E_3 &=&- \Delta \left( m_1 - m_2 - m_3 \right)  \\
 E_3 &=& -\Delta \left(- m_1 + m_2 - m_3 \right) \\ 
 \label{eq:vac3}
 E_4 &=& -\Delta \left(- m_1 - m_2 + m_3 \right) 
\end{eqnarray}
With this in hand we now take a closer look at the remaining extrema in a number of different cases in order to determine the correct aligned vacuum.
\begin{itemize}
\item $\text{sgn} (\det M)=-1$: Let us for simplicity assume that the first and second quark masses are degenerate $m_1=m_2 \equiv m$. Note that the situation for $m_3>0$ and arbitrary $m$ has already been analyzed. Hence we shall take $m_3<0$. The second part of the vacuum conditions can be written as
\begin{eqnarray}\label{eq:vac}
\sin \theta_1 = \sin \theta_2 \ , \qquad \cos \theta_1 + \cos \theta_2 = - \frac{m}{m_3}
\end{eqnarray}
First we note that there only exists a solution provided $2m_3  \leq m \leq -2 m_3$. If this is not satisfied the correct vacuum must be Eq. \ref{eq:vac0} for $m>-2m_3$ and Eq. \ref{eq:vac3} for $m<2m_3$
\begin{eqnarray}
V_{min} &=& 
\left(
\begin{array}{ccc}
1 & 0 & 0 \\
0 & 1 & 0 \\
0 & 0 & 1
\end{array}
\right) \ , \qquad \quad m > -2m_3 \\
V_{min} &=& 
\left(
\begin{array}{ccc}
-1 & 0 & 0 \\
0 & -1 & 0 \\
0 & 0 & 1
\end{array}
\right)\ , \qquad m<2m_3 
\end{eqnarray} 
If on the other hand $2m_3 \leq m \leq -2m_3$ we can find a nontrivial solution to Eq. \ref{eq:vac}. Note that the first condition in Eq. \ref{eq:vac} leads to $\cos \theta_1 = \pm \cos \theta_2$  where only the plus sign will do or else the second condition cannot be satisfied. Hence we arrive at the following nontrivial solution
\begin{eqnarray}
\cos \theta_1 &=& \cos \theta_2 =  - \frac{m}{2m_3} \ , \qquad \text{with} \qquad  E = \Delta \left( \frac{m^2}{2m_3} +m_3 \right)
\end{eqnarray}
One can check that this is the correct vacuum as compared to the other extrema (Eq. \ref{eq:vac0}-\ref{eq:vac3}). Therefore the correct aligned vacuum is
\begin{eqnarray}
V_{min} &=& 
\left(
\begin{array}{ccc}
e^{i\theta_1} & 0 & 0 \\
0 & e^{i \theta_2} & 0 \\
0 & 0 & e^{i \theta_3}
\end{array}
\right) \ , \qquad 2m_3 \leq m \leq -2m_3
\end{eqnarray}
with 
\begin{eqnarray}
\cos \theta_1 = \cos \theta_2 = - \frac{m}{2m_3} \ , \qquad \cos \theta_3 = \frac{m^2}{2m_3^2} -1 
\end{eqnarray}
The vacuum breaks CP invariance spontaneously. This is a quite remarkable situation first observed by Dashen \cite{Dashen} in which we add a small perturbation to a theory that preserves CP and automatically the vacuum is tilted in a direction in which CP breaks spontaneously. It should be noted that Dashen originally observed this phenomenon for the case where all three masses are negative and $2m_3 \leq m <0$. However as is shown here this phase actually extends into the region where $0<m<-2m_3$, i.e. where only the third quark is negative and the first and second quark masses are positive and small compared to the absolute value of the third quark mass.  
\item $m_1=m_2=-m_3>0$: For reasons that will become clear below we will consider this special case. Here the correct aligned vacuum is
\begin{eqnarray} \label{eq:threeflavorlimit}
V_{min} &=& 
\left(
\begin{array}{ccc}
e^{i \frac{1}{3}\pi} & 0&0  \\
0& e^{i \frac{1}{3}\pi} &0 \\
0&0 & e^{-i \frac{2}{3}\pi} \\
\end{array}
\right) \ , \qquad m_1=m_2=-m_3 >0
\end{eqnarray}
\end{itemize}

\subsection{Four Flavors. $N_f=4$.}

The procedure should be clear by now and our method generalizes straightforwardly to four flavors. Consider the perturbation
\begin{eqnarray}
M &=& 
\left(
\begin{array}{cccc}
m_1 & 0 & 0 & 0 \\
0 & m_2 & 0 & 0 \\
0 & 0 & m_3 & 0 \\
0 & 0 & 0 & m_4
\end{array}
\right)
\end{eqnarray}
where all masses are taken to be real such it preserves CP. By identical arguments as in the two and three flavor case the aligned vacuum must consists of four phases that depend on four angles $\theta_1,\theta_2,\theta_3,\theta_4$ and be of the form
\begin{eqnarray}
V_{min} & = & 
\left( 
\begin{array}{cccc}
e^{i\theta_1} & 0 & 0 & 0 \\
0 & e^{i\theta_2} & 0 & 0 \\
0 & 0 & e^{i\theta_3} & 0 \\
0 & 0 & 0 & e^{i\theta_4}
\end{array}
\right) \ , \qquad \theta_1 +\theta_2 + \theta_3 + \theta_4 = 2\pi n
\end{eqnarray}
with
\begin{eqnarray}
 E &=& - \Delta \left( m_1 \cos \theta_1 + m_2 \cos \theta_2 + m_3 \cos \theta_3 + m_4 \cos \theta_4\right)
\end{eqnarray}
The correct vacuum is then found by solving the vacuum (extremum) conditions
\begin{eqnarray}
m_1 \sin \theta_1 = m_2 \sin \theta_2 = m_3 \sin \theta_3 = m_4 \sin \theta_4 \ , \qquad  \theta_1 +\theta_2 + \theta_3 + \theta_4 = 2\pi n
\end{eqnarray}
In the following we will find the aligned vacuum in a variety of different cases.
\begin{itemize}
\item $\text{sgn}\left( \det M \right)=1$: This is the case when all masses simultaneously are either positive, negative or two are positive and the other two are negative. If a quark mass is positive the energy is minimized if the associated cosine factor equals $+1$ and if a quark mass is negative the energy is minimized if the associated cosine factor equals $-1$. Therefore the aligned vacuum is
\begin{eqnarray}
V_{min} &=& 
\left(
\begin{array}{cccc}
\text{sgn}\ m_1 & 0 & 0 & 0 \\
0 & \text{sgn}\ m_2 & 0 & 0 \\
0 & 0 & \text{sgn}\ m_3 & 0 \\
0 & 0 & 0 & \text{sgn}\ m_4
\end{array}
\right) \ , \qquad \text{sgn} \left( \det M \right) = 1
\end{eqnarray}
Note that the same line of argument cannot be given if $\text{sgn}\left( \det M \right)=-1$. Then an odd number of the masses would be negative requiring an odd number of the angles to be multiples of an odd number of $\pi$. However then all the angles would never add up to an even number of $\pi$ as they should. 
\item $\text{sgn}\left(\det M \right)=-1$ and $m_1=m_2=m_3\equiv m$: In this case we need to consider the vacuum conditions carefully similar to the three flavor case with one or three negative masses. Similar to the three flavor case substituting $\theta_4$ the vacuum conditions can be written as
\begin{eqnarray}
\sin \theta_1 &=& \sin \theta_2 = \sin \theta_3 \\
0 &=&  \sin \theta_ 3 \left( m -m_4 +  m_4 \left( \cos \theta_1 \cos \theta_2+ \cos \theta_1 \cos \theta_3+  \cos \theta_2 \cos \theta_3  + \cos^2 \theta_1 \right) \right) 
\end{eqnarray}
They can be satisfied in two different ways. Either
\begin{eqnarray}
\sin \theta_1 &=& \sin \theta_2 = \sin \theta_3 =0 
\end{eqnarray}
or
\begin{eqnarray}\label{eq:fourflavors1}
\sin \theta_1 &=& \sin \theta_2 = \sin \theta_3 \\
\label{eq:fourflavors2}
0 &=& m -m_4 +  m_4 \left( \cos \theta_1 \cos \theta_2+ \cos \theta_1 \cos \theta_3+  \cos \theta_2 \cos \theta_3  + \cos^2 \theta_1 \right)
\end{eqnarray}
The first set of conditions can be satisfied either if all four angles are even multiples of $\pi$, all four angles are odd multiples of $\pi$ or two are even multiples of $\pi$ with the other to being odd multiples of $\pi$. These correspond to the solutions found above. The value of the energy at these locations is
\begin{eqnarray}
\label{eq:vacc1}
E_1 &=& -\Delta (3m +m_4) \\
E_2 &=& -\Delta (-3m -m_4) \\
E_3 &=& -\Delta (m -m_4) \\
\label{eq:vacc2}
E_4 &=& -\Delta (-m +m_4)
\end{eqnarray}
The second set of conditions only has solutions provided $-4 \leq \frac{m_4 - m}{m_4} \leq 4$. If this is not satisfied the vacuum will be one of the four extrema found above. Hence let us assume that the masses are such that it is satisfied. First note that the first set of conditions (Eq. \ref{eq:fourflavors1}) leads to $\cos \theta_1 = \pm \cos \theta_2$ and $\cos \theta_2 = \pm \cos \theta_3$. Here again only the plus signs will do or else the second set of conditions (Eq. \ref{eq:fourflavors2}) cannot be satisfied. We therefore arrive at the following nontrivial solution
\begin{eqnarray}
\cos \theta_1 &=& \cos \theta_2 = \cos \theta_3 =\pm \left( \frac{m_4-m}{4m_4} \right)^{\frac{1}{2}} 
\end{eqnarray}
with
\begin{eqnarray}
E = - \Delta \left( \pm 3 (m-m_4) \left(\frac{m_4-m}{4m_4} \right)^{\frac{1}{2}}  \pm 4 m_4 \left( \frac{m_4-m}{4m_4}\right)^{\frac{3}{2}}\right)
\end{eqnarray}
Note that we do not run into the problem of the square root becoming complex since we have assumed that $m$ and $m_4$ differ in sign. When  $m<0$ and $m_4>0$ the negative sign solution corresponds to the minimum and in the reverse case when $m>0$ and $m_4<0$ the positive sign solution corresponds to the minimum. One can also check that this nontrivial solution is the correct vacuum compared to the other extrema Eq. \ref{eq:vacc1}-\ref{eq:vacc2}. Therefore the aligned vacuum  is
\begin{eqnarray}
V_{min} &=& 
\left( 
\begin{array}{cccc}
e^{i\theta_1} & 0 & 0 & 0 \\
0 & e^{i\theta_2} & 0 & 0 \\
0 & 0 & e^{i\theta_3} & 0 \\
0 & 0 & 0 & e^{i\theta_4}
\end{array}
\right) \ , \qquad -4 < \frac{m_4-m}{m_4} <4
\end{eqnarray}
with 
\begin{eqnarray}
\cos \theta_1 &=& \cos \theta_2 = \cos \theta_3 =\pm  \left( \frac{m_4-m}{4m_4} \right)^{\frac{1}{2}} \ , \qquad \cos \theta_4 = 4 \cos^3 \theta_1 - 3 \cos \theta_1
\end{eqnarray}
The plus sign is for $m>0$ and $m_4<0$ and the negative sign is for $m<0$ and $m_4>0$. Again we see that CP is spontaneously broken by adding a suitable (CP preserving) perturbation. 
\item $m_1=m_2=m_3=-m_4>0$: For reasons that will become clear below we will consider this special case. Here the correct aligned vacuum is
\begin{eqnarray} \label{eq:fourflavorlimit}
V_{min} &=& 
\left(
\begin{array}{cccc}
e^{i \frac{1}{4}\pi} & 0&0 &0 \\
0& e^{i \frac{1}{4}\pi} &0 &0 \\
0&0 & e^{i \frac{1}{4}\pi} &0 \\
0 &0& 0 & e^{-i \frac{3}{4}\pi} 
\end{array}
\right) \ , \qquad m_1=m_2=m_3=-m_4 >0
\end{eqnarray}
\end{itemize}

\subsection{$N_f$ Flavors}

Having performed the analysis of the two, three and four flavor cases we are now ready to make some general remarks for an arbitrary number of flavors $N_f$. First the perturbation we consider is
\begin{eqnarray}
M &=& 
\left(
\begin{array}{ccc}
m_1 & & \\
 & \ddots &      \\
 &    & m_{N_f}
\end{array}
\right)
\end{eqnarray}
The same line of arguments as for the two, three and four flavor case implies that the aligned vacuum must be a function of $N_f$ angles and be of the form
\begin{eqnarray}
V &=& \left(
\begin{array}{ccc}
e^{i\theta_1} & & \\
& \ddots & \\
& & e^{i\theta_{N_f}}
\end{array}
\right) \ , \qquad  \sum_{i=1}^{N_f} \theta_i = 2\pi n
\end{eqnarray}
with the energy 
\begin{eqnarray}
E &= & - \Delta \sum_{i=1}^{N_f} m_1 \cos \theta_i 
\end{eqnarray}
In order to find the aligned vacuum we must minimize the energy over all $N_f$ angles subject to the constraint that they add to $2\pi n$. The vacuum conditions are again a generalization of the two, three and four flavor case and read
\begin{eqnarray}
m_1 \sin \theta_1 &=& \ldots = m_{N_f} \sin \theta_{N_f} \ , \qquad \sum_{i=1}^{N_f} \theta_i = 2\pi n
\end{eqnarray}
Our primary goal is to understand what phases the system might exhibit as one varies the mass parameters with special attention given to the possibility that under the right circumstances CP can be spontaneously broken. By considering the three and four flavor case we first observe that it is a necessary condition for CP to be broken that $\text{sgn} (\det M)=-1$. We expect this to also be the case for arbitrary $N_f$ and shall therefore study the two different situations corresponding to $\text{sgn} (\det M) = \pm 1$.
\begin{itemize}
\item $\text{sgn}\left( \det M \right) =1$: Then there must be an even number of negative masses. Hence the energy is minimized when the cosine factors that are associated with the positive (negative) masses are 1 (-1). Hence the correct aligned vacuum is
\begin{eqnarray}
V_{min} & = & 
\left(
\begin{array}{ccc}
\text{sgn}\ m_1 & & \\
 & \ddots & \\
 & & \text{sgn}\ m_{N_f}
\end{array}
\right)\ , \qquad \text{sgn}\left( \det M \right) =1
\end{eqnarray} 
\item $\text{sgn} (\det M) =-1$: We shall not attempt to solve the system for arbitrary masses but establish the existence of a phase where CP is spontaneously broken at a specific point. In the three and four flavor case it is a necessary condition for CP to be broken that the mass of at least a single flavor becomes negative. Hence we will consider the special situation where $m_1 = \ldots = m_{N_f-1} = - m_{N_f} \equiv m > 0$ and search for a nontrivial solution to the vacuum conditions.

First we observe that the vacuum conditions imply that the first angle $\theta_1$ must be related to the next $N_f-2$ angles $\theta_i,\  i=2,\ldots,N_f-1$ via
\begin{eqnarray}
\theta_1 &=& 2\pi n_i + \theta_i \qquad \text{or} \qquad \theta_1 = \left( 2l_i +1 \right)\pi - \theta_i \ , \qquad i=2,\ldots,N_f-1
\end{eqnarray}
It should be clear that the integers $n_i$ and $l_i$ correspond to equivalent vacua. They all yield the same value of the energy and we can take them to vanish for simplicity. We are therefore left with the two possibilities $\theta_1 = \theta_i$ or $\theta_1 = \pi - \theta_i$ for $i=2,\ldots,N_f-1$.
 
On the other hand the first angle $\theta_1$ is also related to $- \theta_{N_f}$ via the same line of arguments as above. Either $\theta_1 = - \theta_{N_f}$ or $\theta_1 = \pi + \theta_{N_f}$. Since $\theta_{N_f} = 2\pi n - \sum_{i=1}^{N_f-1} \theta_i$ we therefore have
\begin{eqnarray}\label{eq:sum}
\theta_1 &= & \sum_{i=1}^{N_f-1} \theta_i  \qquad \text{or} \qquad \theta_1 = \pi - \sum_{i-1}^{N_f-1} \theta_i
\end{eqnarray}
where again the factor of $2\pi n$ is irrelevant. We will now perform the sum over the $N_f-1$ angles. Assume that $k-1$ angles in the sum are identical to the first angle $\theta_1$. Then there is $N_f-1-k$ angles left in the sum with each being equal to $\pi - \theta_1$. Summing up both contributions then gives 
\begin{eqnarray}
\sum_{i=1}^{N_f-1} &=& k\theta_1 + \left( N_f - 1- k \right)\left( \pi -\theta_1\right) \ , \qquad 1 \le k \le N_f-1
\end{eqnarray}
Inserting this expression for the sum into Eq. \ref{eq:sum} therefore gives us two sets of solutions for the vacuum conditions
\begin{eqnarray}\label{eq:sol1}
\theta_1 &=& \frac{k+1-N_f}{2k - N_f} \pi \ , \qquad E= - \Delta m \left( 2k- N_f\right) \cos \theta_1
\end{eqnarray}
and
\begin{eqnarray}\label{eq:sol2}
\theta_1 &=& \frac{k+2 - N_f}{2k+2-N_f}\pi \ , \qquad E= -\Delta m \left(  2k+2 -N_f \right) \cos \theta_1
\end{eqnarray}
Together they form a complete set of solutions to the vacuum conditions. For a given number of flavors $N_f$ they are parameterized by the integer $1 \le k \le N_f-1$ which counts the number of angles that are identical (up to an even number of $\pi$ which is irrelevant). We are interested in the solutions viewed as functions of $k$ that correspond to the minimum of the energy. 

In Appendix \ref{app:classification} it is shown that the second set of solutions with $k=N_f-1$ correspond to the correct minimum of the energy. Here $\theta_1 = \ldots = \theta_{N_f-1} = \frac{1}{N_f}$. Hence the correct aligned vacuum breaks CP spontaneously and is given by
\begin{eqnarray}
V_{min} &=& 
\left(
\begin{array}{cccc}
e^{i \frac{1}{N_f}\pi} & & & \\
& \ddots & & \\
& & e^{i \frac{1}{N_f}\pi} & \\
& & & e^{-i \frac{N_f-1}{N_f}\pi}
\end{array}
\right) \ , \qquad m_1 = \ldots m_{N_f-1} =- m_{N_f}
\end{eqnarray}
One should note that the results of Eq. \ref{eq:threeflavorlimit} and Eq. \ref{eq:fourflavorlimit} is smoothly recovered in the limit of three and four flavors.

\end{itemize}

\section{Conclusion}\label{sec:conclusion}

We have generalized the original work by Dashen \cite{Dashen} by considering the alignment of the vacuum in fermionic gauge theories with fermions transforming according to a complex representation of the gauge group. The alignment was produced by adding a small mass perturbation. First we considered in detail the situation with four Dirac flavors and then generalized to an arbitrary number of flavors. Here it was shown that at the specific point where $m_1=\ldots=m_{N_f-1}=-m_{N_f}$ CP breaks spontaneously. As one varies the number of flavors this phase is smoothly connected to the Dashen phase for $N_f=3$ and earlier known results are nicely recovered.

Our results show that for CP to be spontaneously broken for any number of flavors it is sufficient that only the mass of a single flavor becomes negative.

\acknowledgments
The CP$^3$-Origins centre is partially funded by the Danish National Research Foundation, grant number DNRF90.

\appendix

\section{Discrete Symmetries}\label{App:Discrete}

The bilinear operators $\bar{\psi}_R \psi_L$ and $\bar{\psi}_L \psi_R$ are of special importance. For the readers convenience we here report how they transform under $C$, $P$ and $CP$
\begin{eqnarray}
\bar{\psi}_{R,f} \psi_L^{f'} & \rightarrow &
\left(
\begin{array}{cc}
\bar{\psi}_{R,f'} \psi_L^{f} \ , & \qquad  \text{under}\ \ C \\
\bar{\psi}_{L,f} \psi_R^{f'} \ , & \qquad \text{under}\ \ P \\
\bar{\psi}_{L,f'} \psi_{R}^{f} \ , & \qquad \text{under}\ \ CP
\end{array}
\right) 
\\
\nonumber
\\
\nonumber
\\
\bar{\psi}_{L,f} \psi_{R}^{f'} & \rightarrow &
\left(
\begin{array}{cc}
\bar{\psi}_{L,f'} \psi_{R}^{f} \ ,  & \qquad \text{under}\ \ C \\
\bar{\psi}_{R,f} \psi_{L}^{f'} \ , & \qquad \text{under}\ \ P \\
\bar{\psi}_{R,f'} \psi_{L}^{f} \ , & \qquad \text{under}\ \ CP
\end{array}
\right)
\end{eqnarray}

\section{Classification of Extrema for Arbitrary $N_f$}\label{app:classification}

Consider the solutions
\begin{eqnarray}
\theta_1 &=& \frac{k+1-N_f}{2k-N_f} \pi \ , \qquad E= -\Delta m \left( 2k-N_f \right) \cos \theta_1 
\end{eqnarray}
for $1 \le k \le N_f-1$. First note that both the coefficient $2k-N_f$ and $\cos \theta_1$ in the expression for the energy simultaneously take their largest positive value when $k=N_f-1$. Here the coefficient is $N_f-2$ and $\cos \theta_1=1$. They also simultaneously take their smallest negative value when $k=1$ for which the coefficient is $-(N_f-2)$ and $\cos \theta_1 =-1$. Hence the energy is minimized for these two solutions with $E=-\Delta m \left(N_f-2 \right)$. 

\begin{figure}[bt]
\centering
\includegraphics[width=0.5\columnwidth]{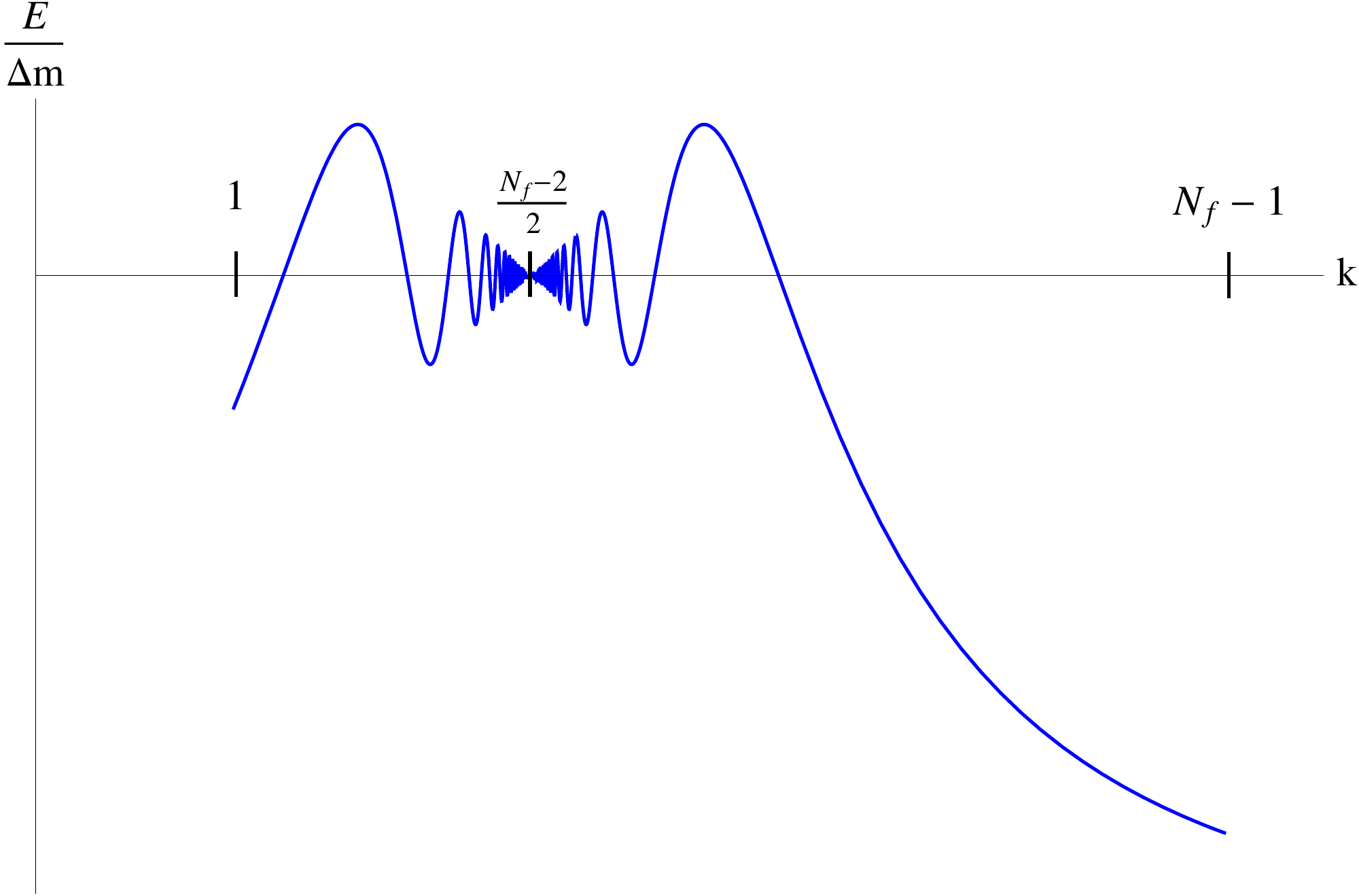}  
 \caption{The energy $E=-\Delta m \left(2k+2-N_f\right) \cos \theta_1,\ \theta_1 =  \frac{k+2-N_f}{2k+2-N_f} \pi$ for a given number of flavors $N_f$. }
\label{plot}
\end{figure}

Next consider the solutions 
\begin{eqnarray}
\theta_1 &=& \frac{k+2-N_f}{2k+2-N_f}\pi \ , \qquad E= - \Delta m \left( 2k + 2 - N_f  \right) \cos \theta_1
\end{eqnarray}
For the ease of argument we assume that $k$ is a continuous parameter. First note that we cannot use the same argument as above since the coefficient $2k+2-N_f$ is again maximized for $k=N_f-1$ but $\cos \theta_1$ is instead maximized for $k=N_f-2$. Instead let us plot the energy as function of $k$ with $1\le k \le N_f-1$ as can be seen in Fig. \ref{plot}. As $k$ approaches the value $\frac{N_f-2}{2}$ the energy oscillates rapidly with the amplitude of oscillation approaching $0$. Then as $k$ increases the amplitude of oscillation similarly increases and the energy reaches it minimum at $k=N_f-1$. Here $E=-\Delta m  N_f \cos \frac{1}{N_f}\pi $. 

For completeness we note that as $N_f$ becomes arbitrarily large the value of the energy at the two endpoints $k=1$ and $k=N_f-1$ coincides.

Finally in order to determine the solution corresponding to the minimum of the energy we need to compare the two values $E= -\Delta m \left( N_f - 2\right) $ and $E= -\Delta m N_f \cos \frac{1}{N_f}\pi$. For $ N_f \ge 3$ we have $\frac{1}{2} \le \cos \frac{1}{N_f}\pi \le 1$ and hence it is easy to check that the second solution 
\begin{eqnarray}
k=N_f-1 \ , \qquad \theta_1 = \frac{1}{N_f}\pi \ , \ \qquad E= -\Delta m N_f \cos \theta_1
\end{eqnarray}
is the correct the minimum of the energy.

\end{document}